\documentclass{article}

\usepackage{amssymb,amsfonts,amsmath}
\usepackage{cite,enumerate,float,indentfirst}
\usepackage{color}

\def\be{\begin{eqnarray}}
\def\ee{\end{eqnarray}}
\def\nn{\nonumber}

\def\i{\mu}
\def\j{\nu}

\hoffset= - 1.0in         

\begin{document}

\hfill ITEP/TH-34/17

\hfill IITP/TH-21/17

\bigskip

\centerline{\Large{
Generalized hypergeometric series for Racah matrices in rectangular representations
}}

\bigskip

\bigskip

\centerline{\bf  A.Morozov }

\bigskip

{\footnotesize
\centerline{{\it
ITEP, Moscow 117218, Russia}}

\centerline{{\it
Institute for Information Transmission Problems, Moscow 127994, Russia
}}

\centerline{{\it
National Research Nuclear University MEPhI, Moscow 115409, Russia
}}
}

\bigskip

\bigskip

\centerline{ABSTRACT}

\bigskip

{\footnotesize
One of spectacular results in mathematical physics is the
expression of Racah matrices for symmetric representations of the quantum group $SU_q(2)$
through the Askey-Wilson polynomials,
associated with the $q$-hypergeometric functions ${_4\phi_3}$.
Recently it was shown that this is in fact the general property of symmetric representations,
valid for arbitrary $SU_q(N)$ -- at least for exclusive Racah matrices $\bar S$.
The natural question then is what substitutes the conventional $q$-hypergeometric polynomials
when representations are more general?
New advances in the theory of matrices $\bar S$, provided by the study of
differential expansions of knot polynomials, suggest that these are multiple sums
over Young sub-diagrams of the one, which describes the original representation of $SU_q(N)$.
A less trivial fact is that the entries of the sum are not just the factorized
combinations of quantum dimensions, as in the ordinary hypergeometric series,
but involve non-factorized quantities, like the skew characters and their further
generalizations -- 
as well as associated additional summations
with the Littlewood-Richardson weights.
}

\bigskip

\bigskip

\section{Introduction}

Racah matrices \cite{racahmatrices} describe the deviation from associativity in the product
of representations and they play a prominent and increasing role in modern
quantum field and string theory.
Despite in many cases we need these quantities for sophisticated representations
of infinite-dimensional algebras, they are still far from being well known even
for the simplest quantum algebras $SU_q(N)$.
In these simple cases Racah matrices are responsible for at least two
subjects of primary importance: modular transformations of conformal blocks \cite{modtrans}
in $2d$ conformal field theory \cite{CFT} and for calculation of knot invariants \cite{knotinvs}
(knot polynomials) in $3d$ Chern-Simons theory \cite{CS}.
Part of the problem is that Racah matrices are actually maps,
and explicit formulas are available in particular basises -- what often
makes these formulas non-invariant (this is often referred to as the
multiplicity problem), and thus not-very-interesting for pure mathematicians.
Thus the progress in the field is largely due to physical methods and
inspirations -- while the rigorous presentation awaits the completion
of the phenomenological part of the story.

In the present paper we open one more new chapter of this exciting mystery-book:
the relation between Racah matrices and hypergeometric orthogonal polynomials
of the Askey-Wilson type \cite{GR}.
Recently we reviewed the subject from the point of view of orthogonal polynomials
\cite{orthopols}, and now we approach it again from the Racah matrix side --
with the natural question: what generalizes hypergeometric series if we switch
from symmetric to generic representations of $SU_q(N)$.

As the first step
we demonstrated in \cite{MMS} that exclusive Racah matrices $\bar S_{\mu\nu}^{R}$,
$$
\Big((R\otimes \bar R)\otimes R \longrightarrow R \Big)
\ \stackrel{\bar S}{\longrightarrow} \
\Big(R\otimes (\bar R \otimes R) \longrightarrow R \Big)
$$
needed for arborescent knot calculus of \cite{arbor},
in the case of symmetric representations $R=[r]$
are expressed through the hypergeometric Racah/Askey-Wilson polynomials
for arbitrary algebra $SU_q(N)$ (for $N=2$ this is the classical result).

Not surprisingly, similar formulas follow from the general
$\bar S$-calculus of {\cite{M3141}--\!\!\!\cite{MnonrectS}}.
Since this calculus is applicable to arbitrary representations $R$,
and is especially well understood for rectangular $R=[r^s]$,
this opens a possibility to generalize the hypergeometric realization.
Relevant substitute of the hypergeometric series in this case are
sums over Young diagrams with the entries made from the skew characters of $su_q(N)$.

We begin from reminding the general ideas of $\bar S$-calculus in sec.\ref{dbfact}
and then describe in sec.\ref{Ffns1} what is currently known about the structure of the
underlying $F$-functions
and their relation \cite{KMtwist} to skew Schur polynomials.
After that in secs.\ref{symreps} and \ref{AW} we explain what happens in
symmetric representations $R=[r]$
and  how the Askey-Wilson realization arises for them in this context.
We conclude in sec.\ref{conc} with the suggestion that the formulas in sec.\ref{dbfact}
are exactly the ones, which provide the extension of Askey-Wilson realization
from symmetric to rectangular representations.
Non-rectangular case is also straightforward, but there are still some technical
difficulties to be resolved to handle non-trivial multiplicities -- the most
interesting part of both Racah and arborescent calculi.
We refer to \cite{Mnonrect,MnonrectS} for explanations and
leave non-rectangular case for the future considerations.

\section{Factorization of differential expansion for double braids
\label{dbfact}
}

According to \cite{M3141,Mfact} and \cite{KMsup,KMtwist} the rectangularly-colored
HOMFLY for a double braid $(m,n)$

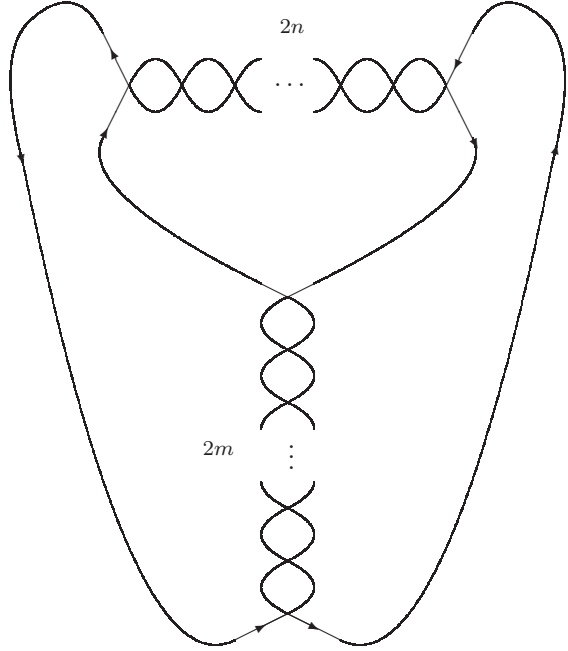
\begin{picture}(200,280)(-230,-230)
\qbezier(-40,0)(-50,20)(-60,0)
\qbezier(-40,0)(-50,-20)(-60,0)
\qbezier(-20,0)(-30,20)(-40,0)
\qbezier(-20,0)(-30,-20)(-40,0)
\qbezier(-20,0)(-15,10)(-10,10)
\qbezier(-20,0)(-15,-10)(-10,-10)
\put(-5,0){\mbox{$\ldots$}}
\qbezier(10,10)(15,10)(20,0)
\qbezier(10,-10)(15,-10)(20,0)
\qbezier(20,0)(30,20)(40,0)
\qbezier(20,0)(30,-20)(40,0)
\qbezier(40,0)(50,20)(60,0)
\qbezier(40,0)(50,-20)(60,0)
\put(-60,0){\line(-1,2){10}}
\put(-60,0){\line(-1,-2){10}}
\put(60,0){\line(1,2){10}}
\put(60,0){\line(1,-2){10}}
\qbezier(0,-80)(-20,-90)(0,-100)
\qbezier(0,-80)(20,-90)(0,-100)
\qbezier(0,-100)(-20,-110)(0,-120)
\qbezier(0,-100)(20,-110)(0,-120)
\qbezier(0,-120)(-10,-125)(-10,-130)
\qbezier(0,-120)(10,-125)(10,-130)
\put(0,-145){\mbox{$\vdots$}}
\qbezier(0,-160)(-10,-155)(-10,-150)
\qbezier(0,-160)(10,-155)(10,-150)
\qbezier(0,-160)(-20,-170)(0,-180)
\qbezier(0,-160)(20,-170)(0,-180)
\qbezier(0,-180)(-20,-190)(0,-200)
\qbezier(0,-180)(20,-190)(0,-200)
\put(0,-80){\line(-2,1){10}}
\put(0,-80){\line(2,1){10}}
\put(0,-200){\line(-2,-1){10}}
\put(0,-200){\line(2,-1){10}}
\put(0,-200){\line(-2,-1){20}}
\put(0,-200){\line(2,-1){20}}
\qbezier(-10,-75)(-80,-40)(-70,-20)
\qbezier(10,-75)(80,-40)(70,-20)
\put(-10,-205){\vector(2,1){2}}
\put(10,-205){\vector(2,-1){2}}
\put(-65,10){\vector(-1,2){2}}
\put(65,10){\vector(-1,-2){2}}
\put(-70,-20){\vector(1,2){2}}
\put(70,-20){\vector(1,-2){2}}
\put(-3,20){\mbox{\footnotesize$2n$}}
\put(-32,-140){\mbox{\footnotesize $2m$}}
%
\qbezier(-70,20)(-80,40)(-97,25)
\qbezier(-97,25)(-111,13))(-100,-30)
\qbezier(-100,-30)(-60,-230)(-20,-210)
\qbezier(70,20)(80,40)(97,25)
\qbezier(97,25)(111,13))(100,-30)
\qbezier(100,-30)(60,-230)(20,-210)
\put(-102,-22){\vector(1,-4){2}}
\put(100,-30){\vector(1,4){2}}
\end{picture}

\noindent
is given by a {\it factorized} differential expansion {\cite{IMMMfe}-\!\!\cite{diffexpan}}:
\be
H_R^{(m,n)}
=  \sum_{\mu,\nu\subset R}
\frac{\sqrt{{\cal D}_\mu{\cal D}_\nu}}{d_R}\,\bar S_{\mu\nu}^{R}\, \Lambda_\mu^m\Lambda_\nu^n
=  \sum_{\lambda\subset R} \chi^*_{\lambda^{tr}}(r)\chi^*_\lambda(s)\cdot
\{q\}^{2|\lambda|} h_\lambda^2\, \chi^*_\lambda(N+r)\chi^*_\lambda(N-s) \cdot
\frac{F_\lambda^{(m)} F_\lambda^{(n)}}{F_\lambda^{(1)}F_\lambda^{(-1)}}
\label{HRdb}
\ee
Here $\{x\} = x-x^{-1}$, quantum numbers are $[n]=\frac{\{q^n\}}{\{q\}}$ and
$h_\lambda$ is denominator in
\be
\chi^*_\lambda(N) =
{\rm Schur}\!\left\{p_k=\frac{[Nk]}{[k]}\right\} =
\frac{1}{h_\lambda} \prod_{(\alpha,\beta)\in \lambda}
[N+\alpha-\beta]
\ee
given by the hook formula:
\be
h_\lambda = \prod_{(\alpha,\beta)\in\lambda} [{\rm hook\ length}_{(\alpha,\beta)}]
= \prod_{(\alpha,\beta)\in\lambda} [{\rm leg}_{(\alpha,\beta)} + {\rm arm}_{(\alpha,\beta)}+1]
\ee
For the figure-eight knot $4_1$ parameters are $(m,n)=(1,-1)$,
and the last factor (the ratio of four $F$) is absent.
Generic twist knots correspond to $n=1$, and, since actually $F_\lambda^{(-1)}=1$
the last ratio turns into just $F^{(m)}_\lambda$.

The difference between knot and Racah calculi
is that for the latter one we need not the $F^{(m)}$ as a total for a given $m$,
but its decomposition into particular eigenvalues, dictated by the evolution method
of \cite{DMMSS} and \cite{evo}.
Among other things this eliminates the factor $\{q\}^{|\lambda|}h_\lambda$ from
the expression for $\bar S$,
making it from just the characters $\chi^*$ at the topological locus \cite{DMMSS}
(quantum dimensions) with a small admixture of additional quantum-number factors.

$F_\lambda^{(m)}$ are actually sums over $m$-the powers of the squared "eigenvalues"
$\Lambda_\mu^{m}$, where $\mu$ are Young sub-diagrams of $\lambda$ --
and according to \cite{Mfact} one can read from this eigenvalue expansion the Racah matrices
$\bar S^{R}_{\mu\nu}$.
According to \cite{KMtwist} the coefficients of the $F$ expansion are essentially
the skew characters $\chi_{\lambda/\mu}=\sum_\nu C^\lambda_{\mu\nu}\chi_\nu$
with the Littlewood-Richardson coefficients $C^\lambda_{\mu\nu}$, defined from
$\chi_\mu\chi_\nu = \sum_\lambda C^\lambda_{\mu\nu} \chi_\lambda$.

Eq.(\ref{HRdb}) and the general formula for $F$-expansion in \cite{KMtwist}
implies  for the case of rectangular representation $R=[r^s]$ with $s\leq 2$ that
\be
\!\!\!
\boxed{
\frac{\sqrt{{\cal D}_\mu{\cal D}_\nu}}{d_R}\,\bar S_{\mu\nu}^{[r^s]}\!
= \!\!\!\!\!\!\!\sum_{\mu,\nu\subset \lambda\subset R}\!\!\!\!\!\!\!\!
\,\frac{(-)^{|\lambda|}
\chi^*_{\lambda^{tr}}(r) \chi^*_\lambda(s) \,\chi^*_\lambda(N+r)\chi^*_\lambda(N-s)}
{\chi^*_\lambda(N)^2}
\!\cdot\! G^\mu_\lambda G^\nu_\lambda\!\! \cdot\!
\frac{\!\chi_{\lambda/\mu}^*(N_\mu^\lambda)\,\chi_{\mu^{tr}}^*(N_\mu^\lambda)}
{\chi_\lambda^*(N_\mu^\lambda)}
\!\cdot\! \frac{\!\chi_{\lambda/\nu}^*(N_\nu^\lambda)\,\chi_{\nu^{tr}}^*(N_\nu^\lambda)}
{\chi_\lambda^*(N_\nu^\lambda)}
}\!\!\!\!
\label{bSexpansion}
\ee
where
$N_\mu^\lambda = N + \sum i - \sum j + \lambda\text{-dependent shift}\ $
for the hook-parametrization \cite{Mfact} of $\mu = (i_1,j_1|i_2,j_2|\ldots )$
and $G_\lambda^\mu$ are some factorized quantities:
ratios of quantum numbers $\ [\,N+u\,]\ $ with various shifts $u$,
also described in terms of the hook parameters $\{i,j\}$ for $\mu$ and $\{a,b\}$ for $\lambda$.
As to ${\cal D}_\mu$ at the l.h.s., they are dimensions of representations,
appearing in decomposition of the product $R\otimes \bar R$ --
which for rectangular $R$ are in one-to-one correspondence with the Young sub-diagrams
of $R$ \cite{KMsup}.

\section{$F$-functions and $G$-factors
\label{Ffns1}}

According to \cite{Mfact} and \cite{KMtwist} the $F$-functions are best described
in a peculiar hook parametrization of Young diagrams:

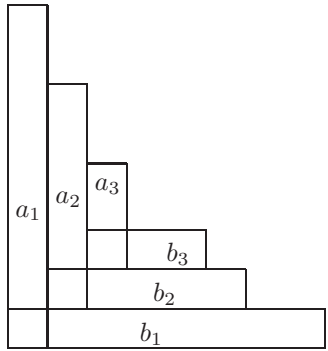
\begin{picture}(300,150)(-50,-10)
\put(0,0){\line(0,1){130}}
\put(15,0){\line(0,1){130}}
\put(30,15){\line(0,1){85}}
\put(45,30){\line(0,1){40}}
\put(75,30){\line(0,1){15}}
\put(90,15){\line(0,1){15}}
\put(120,0){\line(0,1){15}}

\put(0,0){\line(1,0){120}}
\put(0,15){\line(1,0){120}}
\put(15,30){\line(1,0){75}}
\put(30,45){\line(1,0){45}}
\put(0,130){\line(1,0){15}}
\put(15,100){\line(1,0){15}}
\put(30,70){\line(1,0){15}}

\put(3,50){\mbox{$a_1$}}
\put(18,55){\mbox{$a_2$}}
\put(33,60){\mbox{$a_3$}}
\put(50,3){\mbox{$b_1$}}
\put(55,18){\mbox{$b_2$}}
\put(60,33){\mbox{$b_3$}}

\put(130,100){\mbox{a 3-hook Young diagram $\ (a_1,b_1|a_2,b_2|a_3,b_3)$}}
\put(130,80){\mbox{$= \, [a_1+1,a_2+2,a_3+3,3^{b_3},2^{b_2-b_3-1},1^{b_1-b_2-1}]$}}
\put(130,60){\mbox{of the size $\ a_1+a_2+a_3+b_1+b_2+b_3+3$}}

\end{picture}

\noindent
The main drawback of this parametrization is that it changes discontinuously
with the number of hooks:
the empty diagram $\emptyset$ is not a particular case of any 1-hook diagram
$(a_1,b_1)$, of which the minimal is $[1]=(0,0)$ and so on.
Formally one could associate $\emptyset$ with $a_1+b_1=-1$, but this is not
quite respected by the formulas.
Because of this one needs to write
\be
F_\lambda^{(m)}(A,q) = \frac{c_\lambda }{\{q\}^{|\lambda|}h_\lambda\cdot \chi^*_\lambda(N)  }
\sum_{\mu\subset\lambda} f_\lambda^\mu(N,q) \cdot \Lambda_\mu^m
\label{Fvsf}
\ee
with different expressions for different
hook numbers $\#^h_\lambda$ and $\#^h_\mu$.
The $F$-functions depend explicitly on $A=q^N$ and $q$, but mostly are made
from the quantum numbers, involving $N$.
The only exceptions are the squared eigenvalues
\be
\Lambda_\mu = \Lambda_{(i_1,j_1|i_2,j_2|\ldots)} =
\prod_{k=1}^{\#^h_\mu} (A\cdot q^{i_k-j_k})^{2(i_k+j_k+1)}
\ee
and the overall coefficients
\be
c_\lambda = c_{(a_1,b_1|a_2,b_2|\ldots)} =
\prod_{k=1}^{\#^h_\lambda} (A\cdot q^{\frac{a_k-b_k}{2}})^{(a_k+b_k+1)}
\ee
Both, however, drop away from the expression (\ref{HRdb}) for the
Racah matrix $\bar S^R $ -- $\Lambda_\mu$ because $\bar S^R_{\mu\nu}$
are coefficients of the $\Lambda$-expansion and $c_\lambda$ because of
cancellations, dictated by the properties:
\be
F_\lambda^{(-1)}=1, \ \ \ \ \ F_\lambda^{(0)}=\delta_{\lambda,\emptyset}, \ \ \ \ \
F_\lambda^{(1)} = (-)^{\sum_{k}  (a_k+b_k+1)}c_\lambda^2
\label{sumrules}
\ee
which are responsible for the simplicity of the differential expansion
{\cite{IMMMfe}-\!\!\cite{diffexpan}}
at the r.h.s. of  (\ref{HRdb})
for respectively the figure-eight knot $4_1$, unknot and   the trefoil $3_1$.
As already mentioned after (\ref{HRdb}), the factors $\{q\}$ and $h_\lambda$
also drop away from the expressions for HOMFLY polynomials and $\bar S$.

The sum rules (\ref{sumrules}) are non-trivial analogues of the elementary identity
\be
\sum_{\mu\subset \lambda} (-)^{|\mu|}\cdot\chi_{\lambda/\mu}\cdot \chi_{\mu^{tr}}
= \delta_{\lambda,\emptyset}
\label{naivesumrule}
\ee
which follows from the defining property of skew characters,
\be
\sum_{\mu\subset \lambda} \chi_{\lambda/\mu}\{p'_k\}\cdot \chi_{\mu^{tr}}\{p''_k\} =
\chi_\lambda\{p'_k+p''_k\}
\ee
\vspace{-0.6cm}

\noindent
and the transposition law
\vspace{-0.4cm}
\be
\chi_\mu\{-p_k\} = (-)^{|\mu|} \chi_{\mu^{tr}}\{p_k\}
\ee
While (\ref{naivesumrule}) holds beyond the topological locus (i.e. for all values
of time variables), it does not survive introduction of weights $\Lambda^{\pm 1}$
even on the locus,
i.e. there is no analogue of the other two identities in (\ref{sumrules}).

The difficult part of the story is to describe $f_\lambda^\mu$ which satisfy all the three.
Currently they are fully known for $\lambda=(a_1,b_1|a_2,0)$ --
what is enough to get the Racah matrices $\bar S$ for the case $R=[r,r]$
(actually, for this purpose $b_1=0,1$ is sufficient).
After (\ref{naivesumrule}) it is not such a big surprise that they involve skew characters,
but exact formulas \cite{Mfact,KMtwist} are still not very easy to interpret and understand.

$\bullet$ For the empty diagram $\mu$ always
\vspace{-0.3cm}
\be
f_\lambda^\emptyset = 1
\ee

$\bullet$ Since $\mu\subset\lambda$ the number of hooks $\#^h_\mu\leq \#^h_\lambda$.
Thus for the single-hook $\lambda$ it remains to describe only the
contributions of the single-hook $\mu$.
These are relatively simple factorized expressions \cite{Mfact}:
\be
f_{(a,b)}^{(i,j)} = g_{(a,b)}^{(i,j)} \cdot K_{(a,b)}^{(i,j)}
= (-)^{i+j+1}
\cdot \frac{[a]!}{[a-i]![i]!}\cdot\frac{[b]!}{[b-j]![j]!}\cdot\frac{[a+b+1]}{[i+j+1]}
\cdot \frac{D_a!D_i!}{D_{a+i+1}!}\cdot\frac{\bar D_b!\bar D_j!}{\bar D_{b+j+1}!}
\cdot \frac{D_{2i+1}D_{-2j-1}}{D_0D_{i-j}}
\label{f1vsgK}
\ee
with
\vspace{-0.5cm}
\be
g_{(a,b)}^{(i,j)}
= (-)^{i+j+1} \cdot \frac{D_{2i+1}\bar D_{2j+1}}{D_0D_{i-j}}\cdot
\frac{(D_a!)^2}{D_{a+i+1}!\,D_{a-i-1}!}\cdot\frac{(\bar D_b!)^2}{\bar D_{b+j+1}!\,\bar D_{b-j-1}!}
\label{gfactors}
\ee
and
\vspace{-0.5cm}
\be
K_{\lambda}^{\mu}(N) = \frac{\chi^*_{\lambda/\mu}(N)\,\chi^*_{\mu}(N)}{\chi^*_{\lambda}(N)}
\ee
Note that this combination involves $\chi_\mu$ rather than $\chi_{\mu^{tr}}$,
thus $\sum_\mu (-)^{|\mu|}K_{\lambda}^\mu \neq 0$ (in fact, it vanishes, but only for
diagrams $\lambda$ of odd size $|\lambda|=odd$, because
$(-)^{|\mu|}\chi_\mu\{p_k\} = \chi_\mu\{(-)^kp_k\} $, i.e. only odd times change sign).
Notation in (\ref{gfactors}) is: $D_a=[N+a]$, $\bar D_b=[N-b]$ and $D_a!=\prod_{k=0}^a D_k
= \frac{[N+a]!}{[N-1]!}$,
$\bar D_b! = \prod_{k=0}^b \bar D_k = \frac{[N]!}{[N-b-1]!}$
(note that these products start from $k=0$ and include
respectively $a+1$ and $b+1$ factors).

$\bullet$ For two-hook $\lambda=(a_1,b_1|a_2,b_2)$ the formulas are far more involved,
and they are different for different number of hooks in $\mu$:
\be
f_{(a_1,b_1|a_2,b_2)}^{(i_1,j_1)}
= f^{(a_1,b_1)}_{(i_1,j_1)} \cdot \xi_{(a_1,b_1|a_2,b_2)}^{(i_1,j_1)}
= g_{(a_1,b_1)}^{(i_1,j_1)}\cdot K_{(a_1,b_1)}^{(i_1,j_1)}(N)
\cdot \xi_{(a_1,b_1|a_2,b_2)}^{(i_1,j_1)}
\ee
\be
f_{(a_1,b_1|a_2,b_2)}^{(i_1,j_1|i_2,j_2)} =
\frac{[N+i_1+i_2+1][N-j_1-j_2-1]}{[N+i_1-j_2][N+i_2-j_1]}
\cdot \underbrace{g_{(a_1,b_1)}^{(i_1,j_1)}\,g_{(a_2,b_2)}^{(i_2,j_2)}
\cdot K_{(a_1,b_1 )}^{(i_1,j_1 )}(N)
K_{( a_2,b_2)}^{( i_2,j_2)}(N)}_{f_{(a_1,b_1)}^{(i_1,j_1)}\,f_{(a_2,b_2)}^{(i_2,j_2)}}
\cdot\, \xi_{(a_1,b_1|a_2,b_2)}^{(i_1,j_1|i_2,j_2)}
\ee
Non-trivial are the correction factors:
\be
\xi_{(a_1,b_1|a_2,b_2)}^{(i_1,j_1)} =
\  \boxed{\delta_{a_2\cdot b_2,0}}\cdot
\left(
\frac{[N+a_2-j_1][N-b_2+i_1]}{[N+a_2+i_1+1][N-b_2-j_1-1]}
\cdot  {\frac{K_{(a_1,b_1|a_2,b_2)}^{(i_1,j_1)}(N+i_1-j_1) }
{K_{(a_1,b_1)}^{(i_1,j_1)}(N+i_1-j_1)}}
\cdot \boxed{\delta_{i_1\cdot j_1,0}}\ +
\right.
\nn\ee
\vspace{-0.4cm}
\be
\!\!\!\!\!
+ \
\left.
 { \frac{K_{(a_1,b_1|a_2,b_2)}^{(i_1,j_1)}
 \Big(N+(i_1+1)\delta_{b_2,0}-(j_1+1)\delta_{a_2,0}\Big)
 }
{K_{(a_1,b_1 )}^{(i_1,j_1)}\Big(N+(i_1+1)\delta_{b_2,0}-(j_1+1)\delta_{a_2,0}\Big)}
}
\cdot\ \boxed{(1-\delta_{i_1,0})(1-\delta_{j_1,0})}
\right)
+ \ \boxed{(1-\delta_{a_2,0})(1-\delta_{b_2,0}) \ \cdot \, ?}
\label{xi1}
\ee
and
\be
\!\!\!
\xi_{(a_1,b_1|a_2,b_2)}^{(i_1,j_1|i_2,j_2)} =
{\frac{\boxed{\delta_{a_2\cdot b_2,0}}\cdot K_{(a_1,b_1|a_2,b_2)}^{(i_1,j_1|i_2,j_2)}
\Big(N+(i_1+i_2+2)\cdot \delta_{b_2,0} - (j_1+j_2+2)\cdot \delta_{a_2,0}\Big)
}
 {\Big( K_{(a_1,b_1 )}^{(i_1,j_1)}
 \cdot K_{(a_2,b_2 )}^{(i_2,j_2)}\Big)
 \Big(N+(i_1+i_2+2)\cdot \delta_{b_2,0} - (j_1+j_2+2)\cdot \delta_{a_2,0}\Big)}}
 + \ \boxed{(1-\delta_{a_2,0})(1-\delta_{b_2,0}) \ \cdot \, ?}
\nn\ee
Note that we provide expressions only for the case when $a_2\cdot b_2=0$
(i.e. when either $b_2=0$ or $a_2=0$),
what is emphasized by boxes in above formulas.
Sufficient for all the simplest non-symmetric rectangular representations
$R=[r,r]$ and $R=[2^r]$ are respectively $b_2=0$ and $a_2=0$.
Note also, that $K$ factorizes nicely for the single-hook diagrams $\lambda$:
\be
K_{(a,b)}^{(i,j)}(N) =
\frac{\chi^*_{(a,b)/(i,j)}(N)\cdot \chi^*_{(i,j) }(N)}{\chi^*_{(a,b)}(N)} =\nn \\
= \frac{[a]!}{[i]!\,[a-i]!}\cdot\frac{[b]!}{[j]!\,[b-j]!}\cdot \frac{[a+b+1]}{[i+j+1]}
\cdot[N]\cdot \frac{[N+i]!}{[N-j-1]!}\cdot\frac{[N+a-i-1]!}{[N+a]!}
\frac{ [N-b-1]!}{[N-b+j]!}
\label{K1}
\ee
but does not do so
for the two-hook $\lambda$, even if $\mu$ is still a single-hook,
like in (\ref{xi1}), e.g.
\be
K_{(3,1|1,0)}^{(1,1)}(N) \sim  \chi^*_{[43]/[21]}(N)
= \chi^*_{[31]}(N)+\chi^*_{[22]}(N) \sim \nn \\
\sim
A^2q^8+2A^2q^6+A^2q^4+A^2q^2-q^6-q^4-2q^2-1 \sim
[3][2][N+2]+[4][N] \sim
[2]^2[N+2]+[N-2]
\ee
Therefore there seems to be no freedom to change the somewhat mysterious
shifts of $N$ in these formulas.

In the case of $a_2\cdot b_2\neq 0$ skew characters are further deformed
into still more complicated quantities \cite{M333}, which still lack a proper
identification.

\section{Symmetric representations
\label{symreps}}

For symmetric representations $R=[r]$ with $s=1$
we obtain from (\ref{HRdb}), by recursively substituting
(\ref{Fvsf}), (\ref{f1vsgK}), (\ref{gfactors}) and (\ref{K1}):
\be
\bar\sigma_{\mu\nu}^{[r]} =  d_r\cdot
 \frac{[\mu]! [\nu]![N-1]![N-2]!}{[N+\mu-2]![N+\nu-2]!}\,
\cdot\!\!\!\!\sum_{\mu,\nu\leq \lambda\leq r}\!\!\!\!(-)^\lambda
\frac{[\lambda]!}{[r-\lambda]!\,[\lambda-\mu]!\,[\lambda-\nu]!}
\cdot \frac{[N+r+\lambda-1]![N+\lambda-2]!}{[N+\lambda+\mu-1]![N+\lambda+\nu-1]!}
= \nn
\ee
\vspace{-0.5cm}
\be
= \frac{[r]!}{[\mu]![\nu]!}\,
\frac{\chi_r^*(N)^2}{\chi^*_\mu(N)\chi^*_\mu(N-1)\,\chi^*_\nu(N)\chi^*_\nu(N-1)}\cdot
\sum_{\lambda={\rm max}(\mu,\nu)}^{r}
\frac{(-)^\lambda\cdot [\lambda]!}{[r-\lambda]![\lambda-\mu]![\lambda-\nu]!}
\,\frac{ \chi^*_\lambda(N+r)\chi^*_\lambda(N-1)}{\chi^*_\lambda(N+\mu)\chi^*_\lambda(N+\nu)}
\label{symrepser}
\ee
where used are also explicit expression for dimensions
\be
{\cal D}_\mu =  {[N+2\mu-1][N-1]}
\left(\frac{[N+\mu-2]!}{[\mu]!\,[N-1]!}\right)^2
\ee
of representations in
$R\otimes \bar R = [r]\otimes [r^{N-1}] =
\oplus_{\mu=0}^r [r+\mu,r^{N-2},r-\mu]=\oplus_{\mu=0}^r [2\mu,\mu^{N-2}]$
to convert the original $\ \frac{\sqrt{{\cal D}_\mu{\cal D}_\nu}}{d_r}\,\bar S_{\mu\nu}^{[r]}\ $
into $\ \bar\sigma_{\mu\nu}^{[r]} = (-)^{\mu+\nu} \frac{d_r}{\sqrt{{\cal D}_\mu{\cal D}_\nu}}
\,\bar S_{\mu\nu}^{[r]}\ $.

\bigskip

This expression is partly in terms of quantum dimensions $\chi^*_\lambda$ and does contains
neither skew characters, nor shifts.
The only thing which reminds that the skew characters are somewhere behind the scene
is the presence of factorials $\ [\lambda-\mu]!\,[\lambda-\nu]!\ $ in denominator.
Of course, it is possible to make this explicit, by
rewriting (\ref{symrepser}) in the form (\ref{bSexpansion}):
\be
\!\!\!
\bar\sigma_{\mu\nu}^{[r]} =
\frac{1}{\chi^*_\i(N-1)\,\chi^*_\j(N-1)}\ \cdot \!\!\!\!\!\!
\sum_{\lambda={\rm max}(\mu,\nu)}^{r}
\boxed{(-)^\lambda \chi_r^*(N)\chi^*_r(N+\lambda)\cdot
\frac{ [r]!}{[\lambda]![r-\lambda]!} \cdot\frac{[N-1]}{[N+\lambda-1]}}
\cdot\!
\frac{\chi^*_{\lambda-\mu}(N+\mu)\,\chi^*_{\lambda-\nu}(N+\nu)}
{\chi^*_\lambda(N+\mu)\,\chi^*_\lambda(N+\nu)}
\nn
\ee
with relatively simple  shifts
$N_{[\mu]}^{[\lambda]} = N+\mu-1$, which are independent of $\lambda$,
and  $G$-factors
$G_{[\lambda]}^{[\mu]}=(-)^{\mu}\frac{[N+\lambda-1][N+2\mu-1]}{[N+\lambda+\mu-1][N+\mu-1]}$.

\bigskip

{\bf Example of $R=[1]$:}

\bigskip


In we denote the combination in the box in the last formula
through $B_\lambda$, then

\be
\bar\sigma^{[1]}=  \left(\begin{array}{ccc}
\frac{1}{\chi_0(N-1)\chi_0(N-1)}\left(\frac{B_0 \chi_0(N)\chi_0(N)}{\chi_0(N)\chi_0(N)}
+ \frac{B_1 \chi_1(N)\chi_1(N)}{\chi_1(N)\chi_1(N)} \right) &&
\frac{1}{\chi_0(N-1)\chi_1(N-1)}\cdot \frac{B_1 \chi_1(N)\chi_0(N+1)}{\chi_1(N)\chi_1(N+1)}\\ \\
\frac{1}{\chi_1(N-1)\chi_0(N-1)}\cdot \frac{B_1 \chi_0(N+1)\chi_1(N)}{\chi_1(N+1)\chi_1(N)} &&
\frac{1}{\chi_1(N-1)\chi_1(N-1)}\cdot \frac{B_1 \chi_0(N+1)\chi_0(N+1)}{\chi_1(N+1)\chi_1(N+1)}
\end{array}\right) = \nn \\ \nn \\ \nn \\
\!\!\!\!\!\!\!\!\!\!\! =  \left(\begin{array}{ccc}
  B_0+ B_1   &&
\frac{1}{ \chi_1(N-1)}\cdot \frac{B_1  }{ \chi_1(N+1)}\\ \\
\frac{1}{\chi_1(N-1) }\cdot \frac{B_1   }{\chi_1(N+1) } &&
\frac{1}{\chi_1(N-1)\chi_1(N-1)}\cdot \frac{B_1  }{\chi_1(N+1)\chi_1(N+1)}
\end{array} \right) =
\left(\begin{array}{ccc}  1 && -1 \\ \\ -1 &&
-\frac{1}{[N+1] [N-1]}\end{array}\right)
\ \ \ \ \
\ee
with $B_1=- {[N+1][N-1]} $ and $B_0 =  {[N]^2}$.

After multiplication by $\frac{(-)^{i+j}\sqrt{{\cal D}_i{\cal D}_j}}{d_{[r]}}$
with ${\cal D}_0=1$,
${\cal D}_1 = [N+1][N-1]$  and $d_{[1]}=[N]$ this gives the unitary symmetric Racah matrix
\be
\bar S_{[1]} = \frac{1}{[N]} \left(\begin{array}{ccc}
1 & & \sqrt{[N+1][N-1]} \\ \\ \sqrt{[N+1][N-1]} && -1 \end{array}\right)
\ee

\bigskip

{\bf Example of $R=[2]$:}

\bigskip


Similarly, in this case

\bigskip

\centerline{
{\tiny
$
\left(\!\!\!\!\begin{array}{ccc}
\frac{1}{\chi_0(N-1)\chi_0(N-1)}\left(\frac{B_0 \chi_0(N)\chi_0(N)}{\chi_0(N)\chi_0(N)}
+ \frac{B_1 \chi_1(N)\chi_1(N)}{\chi_1(N)\chi_1(N)}
+\frac{B_2 \chi_2(N)\chi_2(N)}{\chi_2(N)\chi_2(N)}\right) &
\frac{1}{\chi_0(N-1)\chi_1(N-1)}\left( \frac{B_1 \chi_1(N)\chi_0(N+1)}{\chi_1(N)\chi_1(N+1)}
+ \frac{B_2 \chi_2(N)\chi_1(N+1)}{\chi_2(N)\chi_2(N+1)}\right)
& \!\!\!\!\!\!
\frac{1}{\chi_0(N-1)\chi_2(N-1)}\cdot \frac{B_2 \chi_2(N)\chi_0(N+2)}{\chi_2(N)\chi_2(N+2)}
\\ \\
\frac{1}{\chi_1(N-1)\chi_0(N-1)}\left(
\frac{B_1 \chi_0(N+1)\chi_1(N)   }{\chi_1(N+1)\chi_1(N)}+
\frac{B_2 \chi_1(N+1)\chi_2(N)}{\chi_2(N+2)\chi_2(N)}\right) &
\!\!\!\!\!\!\!\!\!\!\! \frac{1}{\chi_1(N-1)\chi_1(N-1)}\left(
\frac{B_1 \chi_0(N+1)\chi_0(N+1)   }{\chi_1(N+1)\chi_1(N+1)}+
\frac{B_2 \chi_1(N+1)\chi_1(N)}{\chi_2(N+2)\chi_2(N+1)}\right)&
\!\!\!\! \frac{1}{\chi_1(N-1)\chi_2(N-1)}\cdot
\frac{B_2 \chi_1(N+1)\chi_0(N+2)}{\chi_2(N+2)\chi_2(N+2)}
\\ \\
\frac{1}{\chi_2(N-1)\chi_0(N-1)}\cdot \frac{B_2 \chi_0(N+2)\chi_2(N)}{\chi_2(N+2)\chi_2(N)} &
\frac{1}{\chi_2(N-1)\chi_1(N-1)}\cdot \frac{B_2 \chi_0(N+2)\chi_1(N+1)}{\chi_2(N+2)\chi_2(N+1)}
& \!\!\!\!
\frac{1}{\chi_2(N-1)\chi_2(N-1)}\cdot \frac{B_2 \chi_0(N+2)\chi_0(N+2)}{\chi_2(N+2)\chi_2(N+2)}
\end{array}\!\!\!\!\right) = \!\!
$
}
}

\bigskip

\bigskip

\centerline{
{\footnotesize
$
=  \left(\begin{array}{ccc}
B_0+B_1+B_2 &
\frac{1}{\chi_1(N-1)}\left( \frac{B_1 }{\chi_1(N+1)}
+ \frac{B_2  \chi_1(N+1)}{ \chi_2(N+1)}\right)
&
\frac{1}{\chi_2(N-1)}\cdot \frac{B_2 }{\chi_2(N+2)}
\\ \\
\frac{1}{\chi_1(N-1)}\left(
\frac{B_1    }{\chi_1(N+1)}+
\frac{B_2 \chi_1(N+1)}{\chi_2(N+2)}\right) &
\frac{1}{\chi_1(N-1)\chi_1(N-1)}\left(
\frac{B_1    }{\chi_1(N+1)\chi_1(N+1)}+
\frac{B_2 \chi_1(N+1)\chi_1(N)}{\chi_2(N+2)\chi_2(N+1)}\right)&
\frac{1}{\chi_1(N-1)\chi_2(N-1)}\cdot
\frac{B_2 \chi_1(N+1)}{\chi_2(N+2)\chi_2(N+2)}
\\ \\
\frac{1}{\chi_2(N-1)}\cdot \frac{B_2 }{\chi_2(N+2)} &
\frac{1}{\chi_2(N-1)\chi_1(N-1)}\cdot \frac{B_2 \chi_1(N+1)}{\chi_2(N+2)\chi_2(N+1)}&
\frac{1}{\chi_2(N-1)\chi_2(N-1)}\cdot \frac{B_2 }{\chi_2(N+2)\chi_2(N+2)}
\end{array}\right) =
$
}
}

\bigskip


\be
=  \left(\begin{array}{ccc}
1 & -1 & 1 \\
-1 & \frac{[N+2][N]-[2]^2}{[2][N+2][N-1]}& \frac{[2]}{[N-1][N+2] } \\
1 &\frac{[2]}{[N-1][N+2] } & \frac{[2]^2}{[N-1][N][N+2][N+3]}
\end{array}\right)
\ee

\bigskip

\noindent
with $B_0=\frac{[N]^2[N+1]^2}{[2]^2}$, $B_1 = -\frac{[N+2][N+1]^2[N-1]}{[2]}$ and
$B_2 = \frac{[N-1][N][N+2][N+3]}{[2]^2}$.

\bigskip

Multiplication by $\frac{(-)^{i+j}\sqrt{{\cal D}_i{\cal D}_j}}{d_{[r]}}$
with ${\cal D}_0=1$,
${\cal D}_1 = [N+1][N-1]$, ${\cal D}_2 = \frac{[N+3][N]^2[N-1]}{[2]^2}$
and $d_{[2]}=\frac{[N][N+1]}{[2]}$
provides the unitary symmetric Racah matrix
\be
\bar S_{[2]} = \frac{[2]}{[N+1][N]}\left(\begin{array}{ccc}
1 & \sqrt{[N+1][N-1]} & \frac{[N]\sqrt{[N+3][N-1]}}{[2]} \\
\sqrt{[N+1][N-1]} & [N+1]\frac{[N+2][N]-[2]^2}{[2][N+2]}&
-\frac{[N]}{[N+2]}\sqrt{ {[N+3][N+1]}} \\
\frac{[N]\sqrt{[N+3][N-1]}}{[2]}
&-\frac{[N]}{[N+2]}\sqrt{ {[N+3]}{[N+1]}} &
\frac{[N]}{[N+2]}
\end{array}\right)
\ee

\section{Hypergeometric series
\label{AW}}

Coming back to (\ref{symrepser}),  one can make
a change of summation variable $\lambda =r-k$
to get a factorial $[k]!$ in the denominator.
Then the sum turns into
\be
\bar\sigma_{\mu\nu}^{[r]} =
\frac{[N-1]\,\chi^*_r(N)}{\chi^*_\mu(N-1)\chi^*_\nu(N-1)}
\!\!\!\!\!\!
\sum_{k=0}^{{\rm min}(r-\mu,r-\nu)}
\!\!\!\!\!\!
 \frac{(-)^{r-k} \,[r-k]!}{[k]![r-\mu-k]![r-\nu-k]!}
\frac{[N+2r-1-k]![N+r-2-k]!}{[N+r+\mu-1-k]![N+r+\nu-1-k]!}
\label{symrepser1}
\ee
what is proportional to the $q$-hypergeometric polynomial
\be
 {_4\phi_3} \left(\begin{array}{c}
\mu-r,\ \ \nu-r,\ \ 1-N-r-\mu,\ \ 1-N-r-\nu \\ -r,\ \ 1-2r-N,\ \ 2-r-N
\end{array}\Big|\, z=q^2\right)
\ee
This looks different but is actually equivalent to the result of \cite{MMS}
for the same $\bar\sigma_{\mu\nu}^{[r]}$:
$$
\!\!\!\!
\frac{([\i]![\j]!)^2[r-\i]![r-\j]!\,[N-1]![N-2]!}{[r+\i+N-1]![r+\j+N-1]!}\cdot
d_r \cdot \!\!\!\!\!\!\!
\sum_{k={\rm max}(\i,\j)}^{{\rm min}(r,\i+\j)}\!\!\!\!\!\!\! (-)^k\,
\frac{[k+N+r-1]!}{([k-\i]![k-\j]!)^2[r-k]![\i+\j-k]![\i+\j+N-k-2]!} \sim
$$
\vspace{-0.4cm}
\be
\sim {_4\phi_3} \left(\begin{array}{c}
\i-r,\ \ \i-r,\ \ \j-r,\ \ \j-r \\ 1-2r-N,\ \ \i+\j+1-r,\ \ \i+\j+N-r-1
\end{array}\Big|\,z=q^2\right)
\label{MMSfla}
\ee
Note that (\ref{symrepser}) and (\ref{symrepser1}) contain just five factorials in denominator,
instead of  seven in the first line of (\ref{MMSfla}), which are usual
in the standard formulas for $SU_q(2)$ Racah matrices.

We remind \cite{orthopols} that in the balanced case, i.e. for $\alpha_1+\ldots+\alpha_{p+1}+1 =
\beta_1+\ldots+\beta_p$, hypergeometric series are expressed through quantum numbers and
\be
\!\!\!\!\!\!\!\!\!
{_{p+1}\phi_p}\left(\begin{array}{c}
\alpha_1\ \ldots\ \alpha_{p+1}\\ \beta_1\ \ldots\ \beta_p\end{array}\Big|\,z=q^2\right)
= \sum_n \frac{\chi^*_n(\alpha_1)\ldots \chi^*_n(\alpha_{p+1})}
{\chi^*_n(\beta_1)\ldots\chi^*_n(\beta_p)} = \nn \\
= \sum_n \frac{[\alpha_1+n-1]!\ldots [\alpha_{p+1}+n-1]!\,[\beta_1-1]!\ldots[\beta_p-1]!}
{[\alpha_1-1]!\ldots [\alpha_{p+1}-1]!\,[\beta_1+n-1]!\ldots [\beta_p+n-1]!\,[n]!}
\ \sim \ \sum_k \frac{[-\beta_1-n]!\ldots[-\beta_p-k]!}
{[-\alpha_1-k]!\ldots [-\alpha_{p+1}-k]!\,[k]!}
\ee

\section{Conclusion
\label{conc}}

In this sense (\ref{bSexpansion}) can be considered as the generalization of the
q-hypergeometric polynomials, which is relevant for description of generic Racah matrices,
at least in rectangular representations.
In non-rectangular case some sub-diagrams $\lambda\in R$ appear with non-trivial
multiplicities and contribute additional terms into this expansion,
see \cite{Mnonrect,MnonrectS} for more details.

In general the elements of Racah matrix $\bar S_{\mu\nu}^R$ are expressed through
quantum numbers, but are not factorized -- for two reasons:
because of the sum over sub-diagrams $\lambda\in R$ and because the items in the sum
are made not just from the nicely-factorized quantum dimensions
$\chi^*_\lambda, \chi^*_\mu, \chi^*_\nu$, but also
from the {\it skew} characters $\chi^*_{\lambda/\mu}$
and $\chi^*_{\lambda/\nu}$
and their further generalizations, like (23) and (31) in \cite{M333},
which do {\it not} have this factorization property.
For $R=[r,r]$
one can alternatively represent $\bar S$ as a triple sum with 
the Littlewood-Richardson weights
\be
\bar S_{\mu\nu}^{[r,r]} = \sum_{\mu'\nu'}
\sum_{\mu,\nu\subset \lambda\subset [r,r]} C^\lambda_{\mu\mu'}C^\lambda_{\nu\nu'}
B^{\lambda}_{\mu'\nu'}
\label{SCCB}
\ee
then $B^\lambda_{\mu'\nu'}$ will be factorized combinations of quantum numbers,
but the number of sums seem to grow further for $R=[r^s]$ with $s\geq 3$
\cite{M333}.

Specifics of symmetric representations is that for them the skew characters
$\chi^*_{[\lambda]/[\mu]} = \chi^*_{[\lambda-\mu]}$ {\it do} factorize,
$C^{[\lambda]}_{[\mu],[\mu']} = \delta_{\mu+\mu',\lambda}$
and also the sum over $\lambda$ is a one-fold sum -- what reduces the generic
triple sum (\ref{SCCB}) over Young diagrams to the ordinary $q$-hypergeometric
polynomial, though of a rather complicated Askey-Wilson type ${_4\phi_3}$
with the fixed hypergeometric argument $z=q^2$.


\section*{Acknowledgements}

This work was performed at the Institute for the Information Transmission Problems
with the support from the Russian Science Foundation, Grant No.14-50-00150.

\end{document}